# *In situ* micropillar compression of an anisotropic metal-organic framework single crystal

Zhixin Zeng,[1] Yuan Xiao,[2] Jeffrey M. Wheeler,[2] and Jin-Chong Tan[*,1]

[1]Multifunctional Materials & Composites (MMC) Laboratory, Department of Engineering Science, University of Oxford, Parks Road, Oxford, OX1 3PJ, U.K.

[2]Laboratory for Nanometallurgy, Department of Materials, ETH Zurich, Vladimir-Prelog-Weg 5, HCI G 503, 8093 Zürich, Switzerland.

[*]E-mail: jin-chong.tan@eng.ox.ac.uk

## ABSTRACT

Understanding of the complex mechanical behavior of metal-organic frameworks (MOF) beyond their elastic limit will allow the design of real-world applications in chemical engineering, optoelectronics, energy conversion apparatus, and sensing devices. Through *in situ* compression of micropillars, the uniaxial stress-strain curves of a copper paddlewheel MOF (HKUST-1) were determined along two unique crystallographic directions, namely the (100) and (111) facets. We show strongly anisotropic elastic response where the ratio of the Young's moduli are $E_{(111)} \approx 3.6 \times E_{(100)}$, followed by extensive plastic flows. Likewise, the yield strengths are considerably different, in which $Y_{(111)} \approx 2 \times Y_{(100)}$ because of the underlying framework anisotropy. We measure the fracture toughness using micropillar splitting. While *in situ* tests revealed differential cracking behavior, the resultant toughness values of the two facets are comparable, yielding $K_c \sim 0.5$ MPa$\sqrt{\text{m}}$. This work provides new insights of porous framework ductility at the micron scale and failure by bonds breakage.

(153 words)

Metal-organic frameworks (MOF) are a versatile family of porous solids constructed from organic and inorganic building blocks, yielding either crystalline or amorphous hybrid materials with vastly tunable physical and chemical properties. MOF crystals and MOF-based composites are being developed to target numerous technological applications, ranging from sensors and dielectrics,[1,2] capture and separations,[3] electroluminescence and lighting,[4,5] to drug delivery,[6] mechanical energy absorption,[7] and catalysis.[8,9] Although the chemical and sorption-related properties of MOF compounds have been systematically studied in the past 25 years,[10] the research on their physical properties, especially pertaining to the mechanical behavior of MOFs, is significantly lacking behind.[11-13]

Hitherto, the majority of studies in the field of "MOF mechanics" are focused on the elastic properties of MOF crystals and structural vibrations, employing experimental techniques such as nanoindentation,[11] Terahertz and Raman spectroscopy,[14,15] and Brillouin scattering;[16,17] while computational studies encompass density functional theory (DFT) calculations of single-crystal elastic constants[18-20] and molecular dynamics (MD) of reversible framework deformation.[21,22] Understanding the plastic behavior (beyond the elastic limit)[23] and ultimate fracture[24] of MOFs are central to practical applications for improved mechanical durability and resilience of the resultant devices. With the advancement of focused ion beam (FIB) milling combined with *in situ* micromechanical testing, the plastic deformation of a wide range of materials have been reported in the past decade, exemplars of which include superalloys, gallium arsenide, bones, bulk metallic glasses and amorphous silica.[25-29] More recently, a uniaxial micropillar compression study of MOF glass revealed extensive plasticity of the amorphized $a_g$ZIF-62 monolith, at least on the micrometer length scale.[30] However, to the best of our knowledge the precise characterization of the elastic-to-plastic transition (yielding), large-strain deformation (plastic flow), and fracture toughness (cracking) of MOF single crystals have not yet been investigated.

In this study, Cu$_3$(BTC)$_2$ (BTC = benzene-1,3,5-tricarboxylate), also known as the "HKUST-1" structure, comprising the copper paddlewheel framework has been chosen as a representative 3-D MOF material to study the stress-strain relationship and fracture behavior of a mechanically anisotropic MOF crystal. HKUST-1 crystallizes in the cubic space group ($Fm\bar{3}m$) and exhibits a large surface area in the range of 1,000 m$^2$/g.[31] The ordered nanoporous structure of HKUST-1, its ease of synthesis, and tunable functionalities due to its open metal sites, made this MOF structure one of the most significant in the field. In contrast to an amorphous MOF glass[30] or the polycrystalline monoliths of HKUST-1[32] that are mechanically isotropic,[33] the mechanical response of a single crystal of HKUST-1 is predicted by DFT to be strongly directionally dependent.[34] This gives us the unique opportunity to probe the effects of mechanical anisotropy of a porous MOF, both before and after exceeding the elastic limit and until the point of rupture, by leveraging *in situ* micropillar compression, for the first time.

**Results and discussion**

**Uniaxial compression of micropillars**

*In situ* micro-compression experiments were conducted on the submillimeter-sized HKUST-1 crystals (*ca.* 200−600 $\mu$m), which were prepared using a solvothermal method (see Methods).[35] Activated HKUST-1 crystals were cold-mounted on an epoxy resin (Struers Epofix) with both the (100) and (111) facets exposed (see Figure 1(a)) using an established methodology designed for studying MOF single crystals.[36] The sample surface was meticulously prepared to maximize the surface quality, employing glycerol as the non-penetrating polishing liquid. The large crystal surface was rinsed with ethanol and then desolvated at 80 °C to remove residual liquid. Thereafter, the sample was stored in a desiccator until testing (Supplementary Figure S1).

In order to perform the *in situ* uniaxial compression test, the HKUST-1 crystals were machined into isolated micropillars using gallium focused ion beam (Ga FIB), see Supplementary Figure S2. The relatively large diameters of these MOF pillars were chosen to be ~5 μm for the purpose of mitigating the ion damage (all calculations were based on the precise dimensions of the individual pillar, summarized in Supplementary Table S1). Six HKUST-1 micropillars (three for each crystal facet) were uniaxially compressed, at a constant strain rate of $1\times10^{-3}$ $s^{-1}$. *In situ* micro-compression tests were performed using the Alemnis instrument equipped with a conical indenter, which has a flat apex in order to exert a uniform compressive force. As shown in Figure 1(b-c), the surface of the samples revealed microscopic pores generated by FIB milling, which is inferred to be only superficial damage rather than larger-scale structural failure on account of the consistency of the load vs. displacement curves measured (*vide infra*).

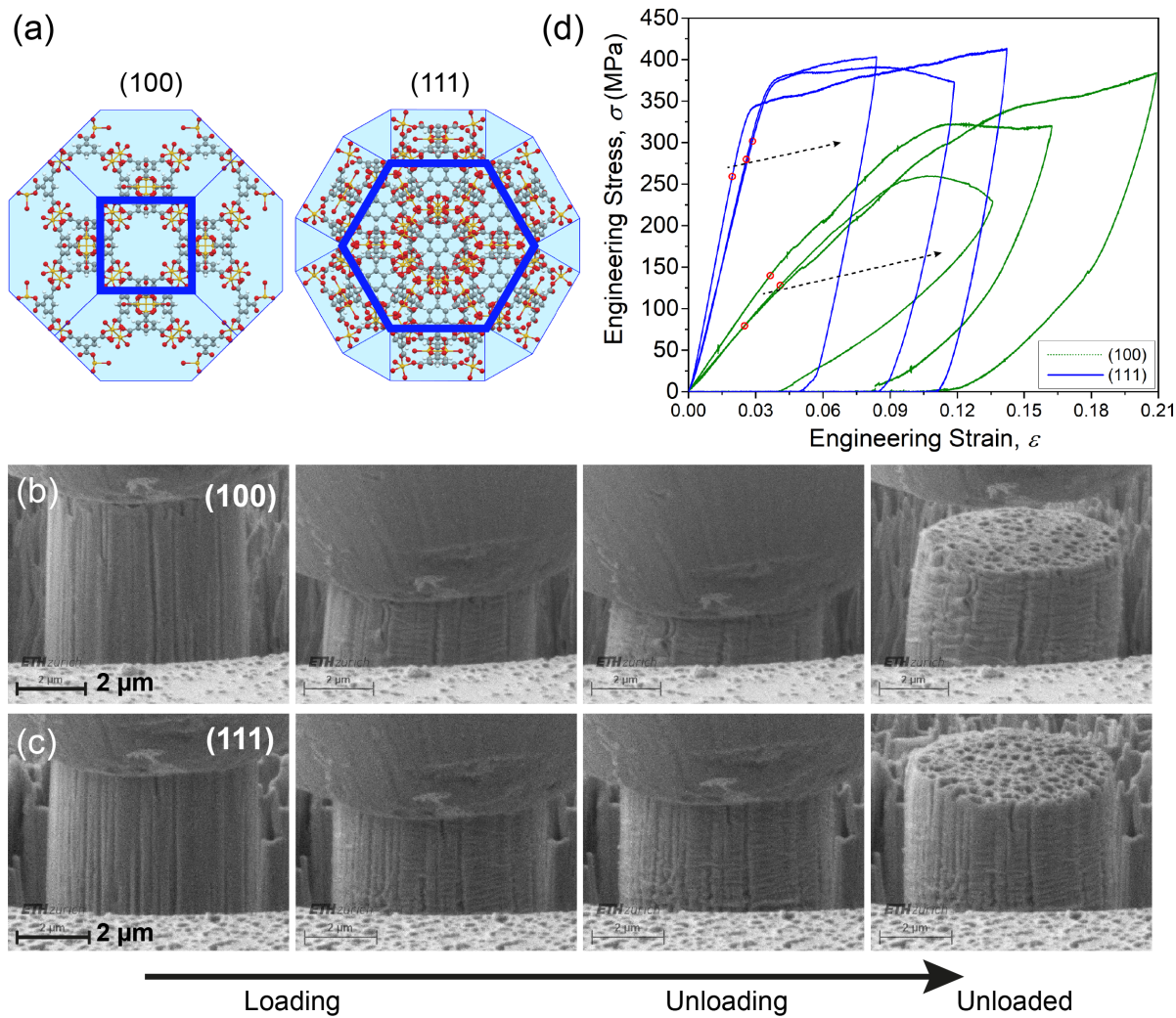

**Figure 1.** (a) Porous crystalline framework of the HKUST-1 MOF structure, viewed down the (100) and (111) crystal facets, respectively. Color code: copper (yellow), oxygen (red), carbon (gray), and hydrogen (white). In the periodic structure, the inorganic copper paddlewheel clusters are bridged by the organic BTC linkers. (b-c) Scanning electron microscope (SEM) images of HKUST-1 micropillars milled by focused ion beam and subsequently compressed uniaxially along the (b) [100]- and (c) [111]-axis, respectively. Recorded videos of the *in situ* compression tests are given in Supplementary Table S2. (d) Engineering stress-strain curves obtained from the compression tests on the (100) and (111) crystal facets, showing their distinctively different elastic-plastic deformation behavior. The yield points (red circles) were established using the first-derivative threshold method (Supplementary Figure S4).

Under uniaxial compression, the (100) facet of HKUST-1 deformed elastically followed by a visible buckling at the base of the pillar (Figure 1(b)). Subsequently, the micropillar sidewall surface experienced micro-buckling and rumpling, which occurred on the layers degraded by the ion beam and this might have contributed to the wider scattering of the force curves for the (100) facet, see Figure 1(d). By contrast, also shown in Figure 1(c, d), the deformation of the (111) facet under the compressive force is considerably more stable. Likewise, there was the elastic deformation before a steady micro-buckling, however, a plastic flow plateau emerged in the stress-strain curves of the (111) facets, in contrast to the apparent non-linear behavior of the (100) facets after an initial average yield point ($Y$) at around 116 MPa.

To examine the stability of the HKUST-1 pillars, we investigated its buckling strength using the Euler's column formula (see Supplementary Section 2). In this column buckling analysis, two possible boundary conditions (BCs) were considered: firstly, the bottom of the pillar was fixed while the top loading surface was free to deflect; secondly, on top of the first scenario, slippage is permitted between the flat punch and the top surface of the compressed pillar. Utilizing the first BC, we determined the Euler's critical load: $P_{E(100)} = 0.85 \pm 0.13$ mN and $P_{E(111)} = 4.51 \pm 1.92$ mN. Conversely, employing the second BC, the magnitudes were elevated to $6.96 \pm 1.06$ mN and $36.99 \pm 15.71$ mN for the (100) and (111) facets, respectively. According to the recorded compression test videos (Supplementary Table S2), it can be seen that the compressed micropillars along the HKUST-1's [100]-axis after unloading had sustained some degree of inclination (hence buckled) as opposed to the pillars along the [111]-axis that remained upright (without buckling) after unloading. This observation supports the buckling strength hypothesis presented above, where $P_{E(111)} > P_{E(100)}$ independent of the imposed BCs.

**Anisotropic elasticity**

The elastic deformation of both facets was reflected by the linear segment of the loading curves. Accordingly, we have measured the Young's moduli ($E$) of the two HKUST-1 facets, employing uniaxial compression: $E_{(100)}$ = 3.44 ± 0.42 GPa and $E_{(111)}$ = 12.37 ± 1.79 GPa. Notably, these stiffness values are in good agreement with the reported theoretical predictions by density functional theory (DFT) calculations of the single-crystal elastic constants of HKUST-1 ($E_{(100)} \approx$ 3 GPa and $E_{(111)} \approx$ 15 GPa).[34] Furthermore, we have used the instrumented nanoindentation technique (IIT) to measure the stiffness anisotropy, employing the MTS Nanoindenter XP instrument equipped with a Berkovich indenter tip (Supplementary Figure S3(a)). In comparison with the results obtained using IIT where the indentation moduli ($M$ used in place of $E$ when the sample's Poisson's ratio, $v_s$ is unknown)[37] were determined as: $M_{(100)}$ = 7.4 ± 0.6 GPa and $M_{(111)}$ = 10.4 ± 0.8 GPa (see Supplementary Figure S3(b)). The uniaxial compression test has excellent agreement with theory, overcoming the averaging effect that was previously affecting the accuracy of IIT when measuring strongly anisotropic crystals. This is a remarkable result, because *via* uniaxial micropillar compression, we have directly measured the anisotropic Young's moduli of HKUST-1 crystal at the highest accuracy thus far attainable. This finding further supports the notion that the influence of the damaged top surface on the modulus measurement was superficial or even negligible, and there was no severe amorphization of the HKUST-1 structure caused by FIB milling.

**Yield strength and plastic flow**

As shown in Figure 1(d), both of the (100) and (111) facets in HKUST-1 develop significant work hardening, to the extent that there is no well-defined yield strength ($Y$). We took the yield stress where it turned into a non-linear portion as the yield strength, *viz.* the proportional limit, even so the yield point may be marginally offset. Despite the linear elastic part of the stress vs. strain ($\sigma-\varepsilon$) curves for the HKUST-1's (111) facet is comparably more discernible, to determine the yield point in a systematically reproducible way, we calculated the first derivative of the stress-strain data (d$\sigma$/d$\varepsilon$) and then utilized the moving average approach to smooth out the first-derivative curve to highlight the fluctuations and trends (Supplementary Figure S4). On this basis, we have determined the yield stress along the [111] crystal axis, which is $Y_{(111)}$ = 280.2 ± 21.4 MPa. Whereas the stress-strain curves obtained from compression test of the (100) facet are not as consistent, resulting in $Y_{(100)}$ = 115.7 ± 24.1 MPa. Hence we observed that $Y_{(111)} \approx 2Y_{(100)}$, which can be rationalized by considering the anisotropic HKUST-1 framework when loaded along the two very dissimilar crystallographic directions, as illustrated in Figure 2(a).

Subsequently, we analyzed the energy absorbed and released in the uniaxial compression test during elastic deformation. The calculated modulus of resilience (i.e., integrated area under $\sigma-\varepsilon$ curve ≤ $Y$) of the HKUST-1's (100) facet equals to $U_{R(100)}$ = 2.63 ± 0.03 J/m$^3$, while it is $U_{R(111)}$ = 3.33 ± 0.84 J/m$^3$ for the (111) facet. Therefore, subject to compression on the (111) crystallographic facet, a unit volume of the HKUST-1 crystal absorbs ~26.6% more strain energy than that of the (100) facet without resulting in plastic deformation.

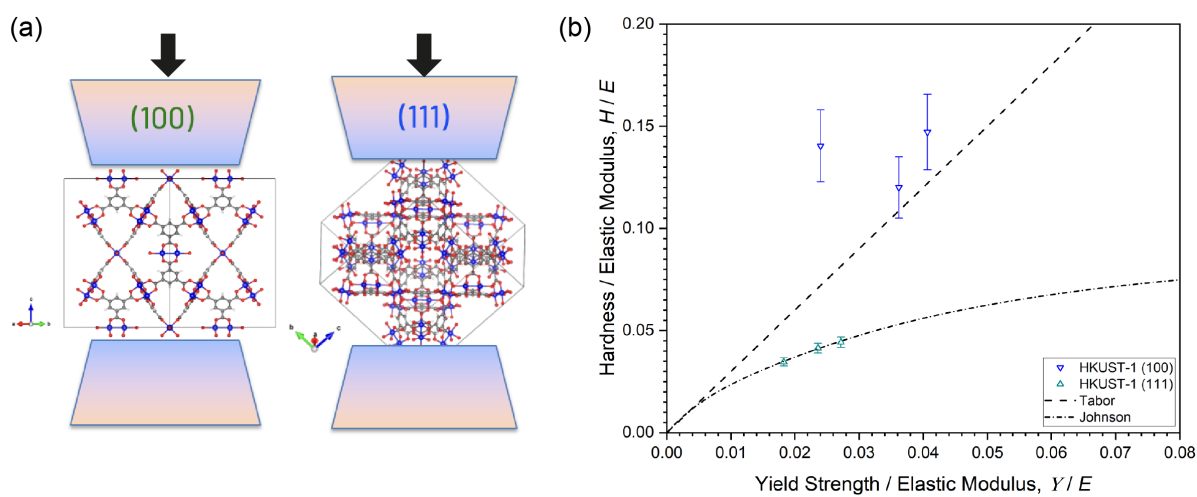

**Figure 2.** (a) Schematics (not to scale) showing the plausible source of mechanical anisotropy associated with the directional effects when compressing the framework structure along the [100]- and [111]-crystal axes. (b) *H/E* vs. *Y/E* for the two HKUST-1 facets in comparison with the Tabor and Johnson correlations.

Unlike the elastic modulus and yield strength values obtained using the micropillar compression test, the values of the hardness (*H*) for the two facets measured by the IIT experiments are quite similar: $H_{(100)} = 463 \pm 58$ MPa and $H_{(111)} = 491 \pm 28$ MPa (see Supplementary Figure S3(c)). During indentation, the material around the indenter is displaced by plastic flow and contributes to its elastic and plastic response. To gain a better understanding of the structural response of the HKUST-1 framework, we studied the correlation between the hardness (*H*) and yield strength (*Y*), when normalized by its Young's modulus (*E*). As can be seen in Figure 2(b), the *H/E* ratio is often correlated with the *Y/E* ratio when the latter is sufficiently large.[38] We found that the *H/E* vs. *Y/E* relation for HKUST-1 (111) is consistent with the result depicted by the cavity theory proposed by Johnson.[39] In this theory, the discrepancy of the volume displaced by the indenter and the elastic expansion is explained by material movement from the plastic deformation zone into the cavity, which is plausible considering the nanoporous nature of MOFs, where framework collapse is expected.[40,41] In general, for different materials, the hardness increases as the yield strength grows due to the reduction of the elastic deformation, which contributes to a smaller contact area. As shown in Figure 2(b), the *H/E* ratio of the (100) facet is approximately 3.4 times greater than that of the (111) facet. The strength of the framework oriented along the [100]- and [111]-axes should differ considerably (Figure 2(a)), which is also reflected by the measured Young's moduli, where $E_{(111)} \approx 3.6 \times E_{(100)}$. Tabor stated that the hardness of rigid plastic materials[42] and metals with work hardening[43] is typically about three times their yield strength ($H \approx 3Y$). However, as established in Figure 2(b), the (100) facet of HKUST-1 exceeds the three-fold relation, which could be ascribed to two possible factors: (i) in Tabor's study, the three-fold correction was used to approximate the mean contact pressure under a spherical indenter for a dense material, but this approximation is not applicable to our case as HKUST-1 has a collapsible nanoporous structure. (ii) As evidenced in Fig. 1(b) (and supplementary videos P1−P3 in Table S2), the (100)-orientation is more susceptible to buckling in its unconfined state, causing its yield strength to be low, while its confined state (during hardness testing) does not permit the framework to buckle following its natural tendency. Consequently, its hardness is even higher than it would be naturally due to its yield strength.

**Fracture toughness measured by micropillar splitting**

For investigating the anisotropic mechanical behavior of HKUST-1 beyond elasticity towards the complete structural failure, a fracture mechanics-based micropillar splitting test was performed along the [100]- and [111]-axes, see Figure 3(a, b). A cube-corner indenter was used to apply the splitting stress starting from the centers of the pillars for both crystal orientations. The abrupt drop in the load in Figure 3(c) corresponds to the rapid and unstable crack propagation. The critical splitting load is indenter geometry-dependent. For instance, the magnitude of the splitting load will be higher if an indenter tip such as a Berkovich tip of larger included angle is employed than that of the sharper cube-corner indenter tip, because of higher stress intensity of the latter. Using the load vs. displacement curves of Figure 3(c), the work done by the splitting load was calculated: $W_{(100)} = 238.2 \pm 45.4$ J/m$^2$; $W_{(111)} = 51.2 \pm 10.7$ J/m$^2$.

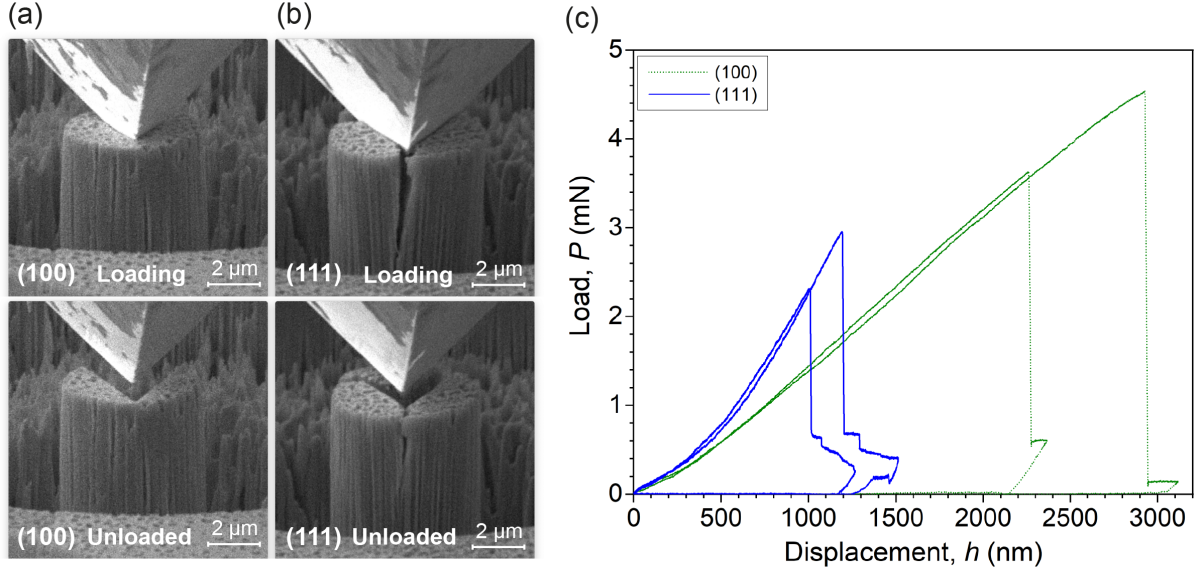

**Figure 3.** *In situ* SEM images of the HKUST-1 micropillars in the splitting tests, when compressed along the (a) [100]- and (b) [111]-oriented axes. Recorded videos of the *in situ* tests are given in Supplementary Table S2. (c) Load vs. displacement (*P-h*) curves acquired from micropillar splitting of the two crystal facets.

Subsequently, we applied the reported outcome of cohesive-zone finite element method[44] to determine the relationship between the stress intensity, the critical load, and the *E/H* ratio for the two facets. The fracture toughness ($K_c$) of the crystals is expressed by:[45]

$$K_c = \gamma \frac{P_c}{R^{3/2}} \quad (1)$$

where $P_c$ represents the critical load at splitting; $R$ denotes the radius of the pillar; and $\gamma$ is the dimensionless coefficient determining the position of the instability within the pillar, which is associated with the *E/H* ratio. $\gamma$ is also temperature-dependent and material-specific elastic-plastic property.[46] Eqn (1) is widely-used for calculating fracture toughness, because the knowledge of crack dimension and sample geometry is not essential. This is a significant advantage for nanomaterials, especially MOFs; where the crack dimension can be difficult to measure, and the propagation of cracks is usually less stable. It is worth noting that a material under a non-plane strain condition usually experiences larger-scale plastic deformation, and this means that $K_c$ is not independent of the size of cracks or defects in the HKUST-1 crystals and the geometry of the specimen. Consequently, the magnitude of the fracture toughness measured here can be considered only as the upper-bound compared with possible other techniques such as single-edge pre-cracked beam method, reported for an isotropic MOF glass of ZIF-62.[47] Herein, we applied the relationship between $\gamma$ and $E/H$ (Supplementary Figure S5) to determine the averaged values of $\gamma$ for both the (100) and (111) facets: 0.543 and 0.828, respectively. As a result, we calculated the fracture toughness: $K_{c(100)} = 0.524 \pm 0.033$ MPa$\sqrt{m}$, which is resembling the magnitude of the other facet, $K_{c(111)} = 0.515 \pm 0.066$ MPa$\sqrt{m}$ in spite of their rather different behavior in terms of the $E/H$ ratio and the load vs. displacement curves.

HKUST-1 crystals are predicted to be highly anisotropic from previous DFT calculations (Zener anisotropy ~5.41 using the B3LYP functional).[34] The similarity of the measured fracture toughness of the two facets is reasoned by the fact that the pillar splitting method imposes three-fold symmetry upon the crack propagation, which averaged out the effect of structural anisotropy. Furthermore, the measured critical stress intensities have been shown to be largely independent of displacement rate effects, within the range used in these measurements.[46] Unlike the compression test, the damage induced by the FIB milling may not be negligible here since the splitting test of silicon pillars was reported to significantly increase the apparent fracture toughness at small pillar sizes, although the influence diminishes to negligibility at the pillar diameters larger than 10 μm.[46] Therefore, further experiment will be required to investigate the influence of the pillar diameter on the fracture toughness of HKUST-1 pillars, although a convergence of the obtained toughness value is expected to be observed.

**Comparison of fracture toughness values**

To consider these results in a broader material's context, the fracture toughness of the single crystal of HKUST-1 can be presented in an Ashby-style plot of $K_c$ versus $E$, as presented in Figure 4. Thus far, there are only a limited number of studies which have characterized the fracture toughness of MOFs and inorganic-organic framework materials. Their values are by and large clustered near to the projected lower-bound for $K_c$ while occupying a gap unpopulated by conventional materials. Of note, the toughness of HKUST-1 crystal is remarkably higher compared with the amorphous ZIF-62 glass ($K_c$ = 0.104 MPa$\sqrt{m}$),[47] as well as against the nanostructured polycrystalline monoliths of ZIF-8 ($K_c$ = 0.074 MPa$\sqrt{m}$) and ZIF-71 ($K_c$ = 0.145 MPa$\sqrt{m}$).[24] As these values were obtained using different geometries, the comparison is not direct, but the scale of the difference implies that significantly higher toughness is likely in the HKUST-1 MOF.

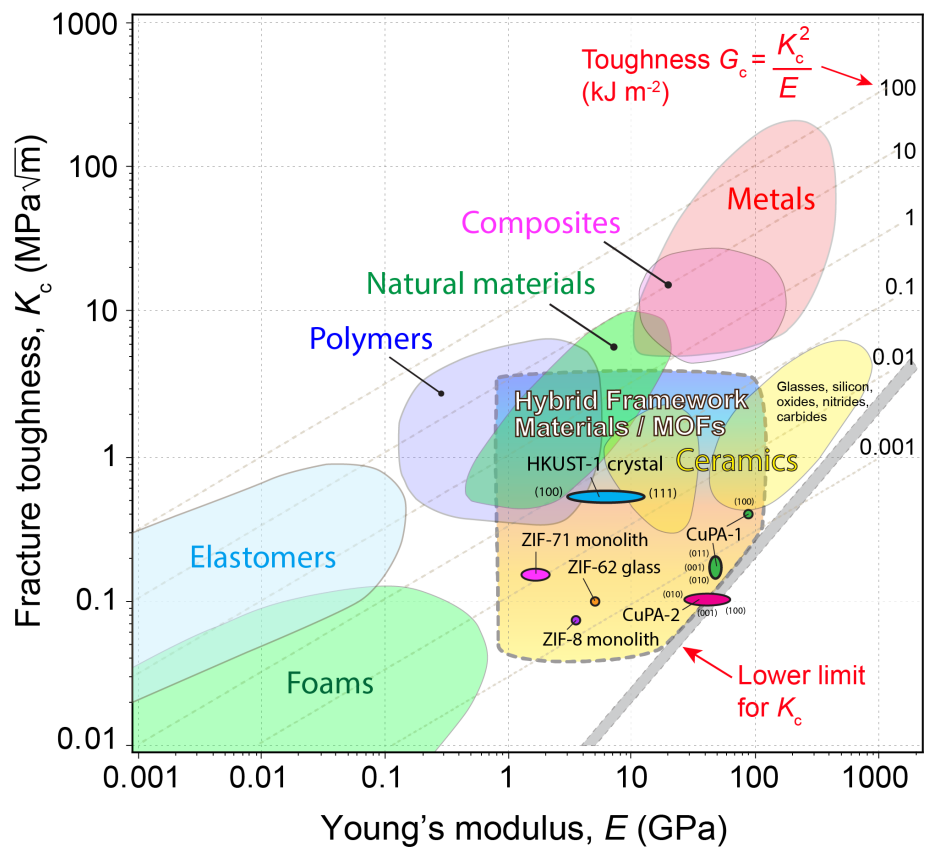

**Figure 4.** Ashby-style plot of fracture toughness ($K_c$) versus Young's modulus ($E$) of HKUST-1 single crystal against common engineering materials and hybrid framework materials, including dense inorganic-organic frameworks,[37] nanocrystalline MOF monoliths (ZIF-8 and ZIF-71),[24] and ZIF-62 glass.[47] The fracture toughness plotted for HKUST-1 crystal represents the upper-bound value using the micropillar splitting technique.

In summary, we have conducted the uniaxial micro-compression, instrumented nanoindentation, and micropillar splitting tests on large (sub-mm) single crystals of the copper paddlewheel MOF, HKUST-1, oriented along the [100]- and [111]-axes. Based on the *in situ* recorded stress-strain curves and fracture history, we have established the anisotropic elastic moduli, yield strength, plastic flow, and fracture toughness of the porous HKUST-1 structure in two distinct crystallographic directions. Moreover, the experimental results of the anisotropic Young's moduli obtained by uniaxial compression demonstrate the veracity of the theoretical values from previous DFT calculations. Insights gained from the plastic flow (a sign of ductility) and fracture behavior of HKUST-1 should instigate exciting research to uncover the large-strain deformation mechanism(s) of porous anisotropic frameworks and inorganic-organic structures. Henceforth, the methodologies we demonstrated herein can be straightforwardly adopted to probe the non-linear mechanics of numerous other MOF crystals beyond elasticity, including the monoliths of covalent organic frameworks (COFs) and hybrid glasses, amongst others.

## METHODS

**Synthesis of large HKUST-1 single crystals.** The solvothermal synthesis of HKUST-1 ($Cu_3BTC_2$, BTC = benzene-1,3,5,-tricarboxylate) uses glacial acetic acid as a modulator since it is effective and reproducible to yield "large" (sub-mm) crystals of about 200−600 μm (Supplementary **Error! Reference source not found.**). Firstly, 0.49 g of $Cu(NO_3)_2·3H_2O$ was fully dissolved in 3 mL of deionized water before combined with 3 mL of *N, N'*-dimethylformamide (DMF) in a scintillation vial (20 mL), forming the $Cu(NO_3)_2$ solution. Meanwhile, we dissolved 0.24 g of trimesic acid ($H_3BTC$) in 3 mL of ethanol (mild heating to assist with dissolution). Thereafter, the $H_3BTC$ solution and 12 mL of glacial acetic acid were successively added to the Cu(NO$_3$)$_2$ solution. The vial containing the mixture of solutions were sealed and heated in oven at 55 °C for 3 days, the reaction yielded blue crystals of HKUST-1 forming on both the wall and the bottom of the vial. The mother liquor was then removed, and the crystals were immersed in ethanol for at least 3 days for solvent exchange. It is worth noting that crystals collected from the bottom of the vial were susceptible to growth defects, since they were prone to form defects compared with crystals attached on the wall of the glass vial. And thus, it is advisable to store the crystals harvested from the wall and the bottom separately in different vials of ethanol. If needed, the crystals can later be activated in a vacuum oven at 120 °C overnight.

**Micropillar compression and micropillar splitting tests.** Sample micro-geometries were produced using a Helios 600i (Thermo Fisher Scientific) FIB workstation operated at an accelerating voltage of 30 keV using a Ga$^+$ focused ion beam. For *in situ* tests, pillars were fabricated with a target diameter of ~5 μm and an aspect ratio of ~3. A two-step milling method was employed with milling currents of 0.79 nA for coarse milling and 40 pA for fine polishing. Micropillar compression testing was performed using an *in situ* nanomechanical testing system (Alemnis AG, Thun, Switzerland) inside a Vega 3 (Tescan, Brno, Czech Republic) SEM. An electrically-conductive diamond flat punch tip with a 8 μm diameter was used to optimize imaging quality. Micropillars were compressed at a constant displacement rate appropriate to generate a strain rate of $1 \times 10^{-3}$ s$^{-1}$. Engineering stress-strain curves were determined from load-displacement curves, using the top diameter of the pillars after correcting for pillar sink-in and instrument compliance.[1] Pillar splitting tests were carried out using the same testing system as used for the micropillar compression tests. A diamond cube corner indenter was used to perform the splitting at a speed of 15 nm/s under displacement control.

**Acknowledgements**

J.C.T. thank the EPSRC award (EP/R511742/1) and ERC consolidator Grant (PROMOFS grant agreement 771575) for funding the research. The authors would like to thank C. Zaubitzer (ScopeM, ETH Zürich) for assistance with FIB of the samples.

**Author contributions**

J.C.T. and J.M.W. conceived the project. Z.X.Z. prepared the crystal samples, performed data analysis, and written first draft of the manuscript under the supervision of J.C.T. Y.X. performed the FIB of micropillars and conducted *in situ* compression experiments under the supervision of J.M.W. All authors discussed the results and contributed to the final version of the manuscript.

**Competing interests**

The authors declare no competing financial interest.

**Additional information**

Supplementary information is available for this paper at {publisher's URL}.

**Author information**

Corresponding Author's e-mail: jin-chong.tan@eng.ox.ac.uk

ORCID

Jin-Chong Tan: 0000-0002-5770-408X

Jeffrey Wheeler: 0000-0001-7763-2610

Zhixin Zeng: 0000-0002-7871-3970

Yuan Xiao: N/A

**References**

1 Gutiérrez, M., Zhang, Y. & Tan, J. C. Confinement of Luminescent Guests in Metal-Organic Frameworks: Understanding Pathways from Synthesis and Multimodal Characterization to Potential Applications of LG@MOF Systems. *Chem. Rev.* **122**, 10438-10483 (2022).

2 Mendiratta, S., Usman, M. & Lu, K.-L. Expanding the dimensions of metal–organic framework research towards dielectrics. *Coord. Chem. Rev.* **360**, 77-91 (2018).

3 Bao, Z. B. *et al.* Potential of microporous metal-organic frameworks for separation of hydrocarbon mixtures. *Energy Environ. Sci.* **9**, 3612-3641 (2016).

4 Gutiérrez, M. *et al.* Highly luminescent silver-based MOFs: Scalable eco-friendly synthesis paving the way for photonics sensors and electroluminescent devices. *Appl. Mater. Today* **21**, 100817 (2020).

5 Lustig, W. P. & Li, J. Luminescent metal-organic frameworks and coordination polymers as alternative phosphors for energy efficient lighting devices. *Coord. Chem. Rev.* **373**, 116-147 (2018).

6 Cai, W., Chu, C. C., Liu, G. & Wang, Y. X. Metal-Organic Framework-Based Nanomedicine Platforms for Drug Delivery and Molecular Imaging. *Small* **11**, 4806-4822 (2015).

7 Sun, Y. *et al.* High-rate nanofluidic energy absorption in porous zeolitic frameworks. *Nat. Mater.* **20**, 1015-1023 (2021).

8 Bavykina, A. *et al.* Metal-Organic Frameworks in Heterogeneous Catalysis: Recent Progress, New Trends, and Future Perspectives. *Chem. Rev.* **120**, 8468-8535 (2020).

9 Ryder, M. R. & Tan, J. C. Nanoporous Metal-Organic Framework Materials for Smart Applications. *Mater. Sci. Tech.* **30**, 1598-1612 (2014).

10 Freund, R. *et al.* 25 Years of Reticular Chemistry. *Angew. Chem. Int. Ed.* **60**, 23946-23974 (2021).

11 Tan, J. C. & Cheetham, A. K. Mechanical Properties of Hybrid Inorganic-Organic Framework Materials: Establishing Fundamental Structure-Property Relationships. *Chem. Soc. Rev.* **40**, 1059-1080 (2011).

12 Li, W., Henke, S. & Cheetham, A. K. Research Update: Mechanical properties of metal-organic frameworks – Influence of structure and chemical bonding. *APL Mater.* **2**, 123902 (2014).

13 Burtch, N. C., Heinen, J., Bennett, T. D., Dubbeldam, D. & Allendorf, M. D. Mechanical Properties in Metal-Organic Frameworks: Emerging Opportunities and Challenges for Device Functionality and Technological Applications. *Adv. Mater.*, 1704124 (2017).

14   Ryder, M. R. *et al.* Identifying the Role of Terahertz Vibrations in Metal-Organic Frameworks: From Gate-Opening Phenomenon to Shear-Driven Structural Destabilization. *Phys. Rev. Lett.* **113**, 215502 (2014).

15   Krylov, A. S. *et al.* Raman spectroscopy studies of the terahertz vibrational modes of DUT-8 (Ni) metal-organic framework. *Phys. Chem. Chem. Phys.* **19**, 32099-32104 (2017).

16   Tan, J. C. *et al.* Exceptionally Low Shear Modulus in a Prototypical Imidazole-Based Metal-Organic Framework. *Phys. Rev. Lett.* **108**, 095502 (2012).

17   Radhakrishnan, D. & Narayana, C. Guest dependent Brillouin and Raman scattering studies of zeolitic imidazolate framework-8 (ZIF-8) under external pressure. *J. Chem. Phys.* **144**, 134704 (2016).

18   Ryder, M. R., Civalleri, B. & Tan, J. C. Isoreticular zirconium-based metal-organic frameworks: discovering mechanical trends and elastic anomalies controlling chemical structure stability. *Phys. Chem. Chem. Phys.* **18**, 9079-9087 (2016).

19   Wang, M., Zhang, X., Chen, Y. & Li, D. How Guest Molecules Stabilize the Narrow Pore Phase of Soft Porous Crystals: Structural and Mechanical Properties of MIL-53(Al)⊃$H_2O$. *J. Phys. Chem. C* **120**, 5059-5066 (2016).

20   Wu, H., Yildirim, T. & Zhou, W. Exceptional Mechanical Stability of Highly Porous Zirconium Metal-Organic Framework UiO-66 and Its Important Implications. *J. Phys. Chem. Lett.* **4**, 925-930 (2013).

21   Tafipolsky, M., Amirjalayer, S. & Schmid, R. First-principles-derived force field for copper paddle-wheel-based metal-organic frameworks. *J. Phys. Chem. C* **114**, 14402-14409 (2010).

22   Ma, Q. *et al.* Guest-modulation of the mechanical properties of flexible porous metal–organic frameworks. *J. Mater. Chem. A* **2**, 9691-9698 (2014).

23   Zeng, Z., Flyagina, I. S. & Tan, J.-C. Nanomechanical behavior and interfacial deformation beyond the elastic limit in 2D metal–organic framework nanosheets. *Nanoscale Adv.* **2**, 5181-5191 (2020).

24   Tricarico, M. & Tan, J.-C. Mechanical properties and nanostructure of monolithic zeolitic imidazolate frameworks: a nanoindentation, nanospectroscopy, and finite element study. *Mater. Today Nano* **17**, 100166 (2022).

25   Michler, J., Wasmer, K., Meier, S., Östlund, F. & Leifer, K. Plastic deformation of gallium arsenide micropillars under uniaxial compression at room temperature. *Appl. Phys. Lett.* **90**, 043123 (2007).

26   Girault, B., Schneider, A. S., Frick, C. P. & Arzt, E. Strength Effects in Micropillars of a Dispersion Strengthened Superalloy. *Adv. Eng. Mater.* **12**, 385-388 (2010).

27   Lacroix, R. *et al.* Micropillar Testing of Amorphous Silica. *International Journal of Applied Glass Science* **3**, 36-43 (2012).

28  Wheeler, J. M., Raghavan, R. & Michler, J. In situ SEM indentation of a Zr-based bulk metallic glass at elevated temperatures. *Materials Science and Engineering: A* **528**, 8750-8756 (2011).

29  Schwiedrzik, J. *et al.* In situ micropillar compression reveals superior strength and ductility but an absence of damage in lamellar bone. *Nat. Mater.* **13**, 740-747 (2014).

30  Widmer, R. N., Bumstead, A. M., Jain, M., Bennett, T. D. & Michler, J. Plasticity of Metal-Organic Framework Glasses. *J. Am. Chem. Soc.* **143**, 20717-20724 (2021).

31  Chui, S. S., Lo, S. M., Charmant, J. P., Orpen, A. G. & Williams, I. D. A chemically functionalizable nanoporous material [$Cu_3(TMA)_2(H_2O)_3$]$_n$. *Science* **283**, 1148-1150 (1999).

32  Bennett, T. D. *et al.* Structure and properties of an amorphous metal-organic framework. *Phys. Rev. Lett.* **104**, 115503 (2010).

33  Tian, T. *et al.* A sol-gel monolithic metal-organic framework with enhanced methane uptake. *Nat. Mater.* **17**, 174-179 (2018).

34  Ryder, M. R., Civalleri, B., Cinque, G. & Tan, J. C. Discovering connections between terahertz vibrations and elasticity underpinning the collective dynamics of the HKUST-1 metal–organic framework. *CrystEngComm* **18**, 4303-4312 (2016).

35  Tovar, T. M. *et al.* Diffusion of CO2 in Large Crystals of Cu-BTC MOF. *J. Am. Chem. Soc.* (2016).

36  Tan, J. C., Bennett, T. D. & Cheetham, A. K. Chemical structure, network topology, and porosity effects on the mechanical properties of Zeolitic Imidazolate Frameworks. *Proc. Natl. Acad. Sci. USA* **107**, 9938-9943 (2010).

37  Tan, J. C., Merrill, C. A., Orton, J. B. & Cheetham, A. K. Anisotropic Mechanical Properties of Polymorphic Hybrid Inorganic-Organic Framework Materials with Different Dimensionalities. *Acta. Mater.* **57**, 3481-3496 (2009).

38  Cheng, Y. T. & Li, Z. Y. Hardness obtained from conical indentations with various cone angles. *J. Mater. Res.* **15**, 2830-2835 (2000).

39  Johnson, K. L. The Correlation of Indentation Experiments. *J. Mech. Phys. Solids* **18**, 115-126 (1970).

40  Banlusan, K., Antillon, E. & Strachan, A. Mechanisms of Plastic Deformation of Metal-Organic Framework-5. *J. Phys. Chem. C* **119**, 25845-25852 (2015).

41  Vandenhaute, S., Rogge, S. M. J. & Van Speybroeck, V. Large-Scale Molecular Dynamics Simulations Reveal New Insights Into the Phase Transition Mechanisms in MIL-53(Al). *Frontiers in Chemistry* **9** (2021).

42  Tabor, D. The physical meaning of indentation and scratch hardness. *Br. J. Appl. Phys.* **7**, 159-166 (1956).

43  Tabor, D. *The Hardness of Metals*. (Clarendon Press, Oxford 1951).

44  Ghidelli, M., Sebastiani, M., Johanns, K. E. & Pharr, G. M. Effects of indenter angle on micro-scale fracture toughness measurement by pillar splitting. *J. Am. Ceram. Soc.* **100**, 5731-5738 (2017).

45  Sebastiani, M., Johanns, K. E., Herbert, E. G., Carassiti, F. & Pharr, G. M. A novel pillar indentation splitting test for measuring fracture toughness of thin ceramic coatings. *Philos. Mag.* **95**, 1928-1944 (2014).

46  Lauener, C. M. *et al.* Fracture of Silicon: Influence of rate, positioning accuracy, FIB machining, and elevated temperatures on toughness measured by pillar indentation splitting. *Mater. Des.* **142**, 340-349 (2018).

47  To, T. *et al.* Fracture toughness of a metal-organic framework glass. *Nat. Commun.* **11**, 2593 (2020).

*Supplementary Information*

*for*

*In situ* **micropillar compression of an anisotropic metal-organic framework single crystal**

Zhixin Zeng,[1] Yuan Xiao,[2] Jeffrey M. Wheeler,[2] and Jin-Chong Tan[*,1]

[1]Multifunctional Materials & Composites (MMC) Laboratory, Department of Engineering Science, University of Oxford, Parks Road, Oxford, OX1 3PJ, U.K.

[2]Laboratory for Nanometallurgy, Department of Materials, ETH Zurich, Vladimir-Prelog-Weg 5, HCI G 503, 8093 Zürich, Switzerland.

[*]E-mail: jin-chong.tan@eng.ox.ac.uk

# Table of Contents

1. **Materials and Methods**

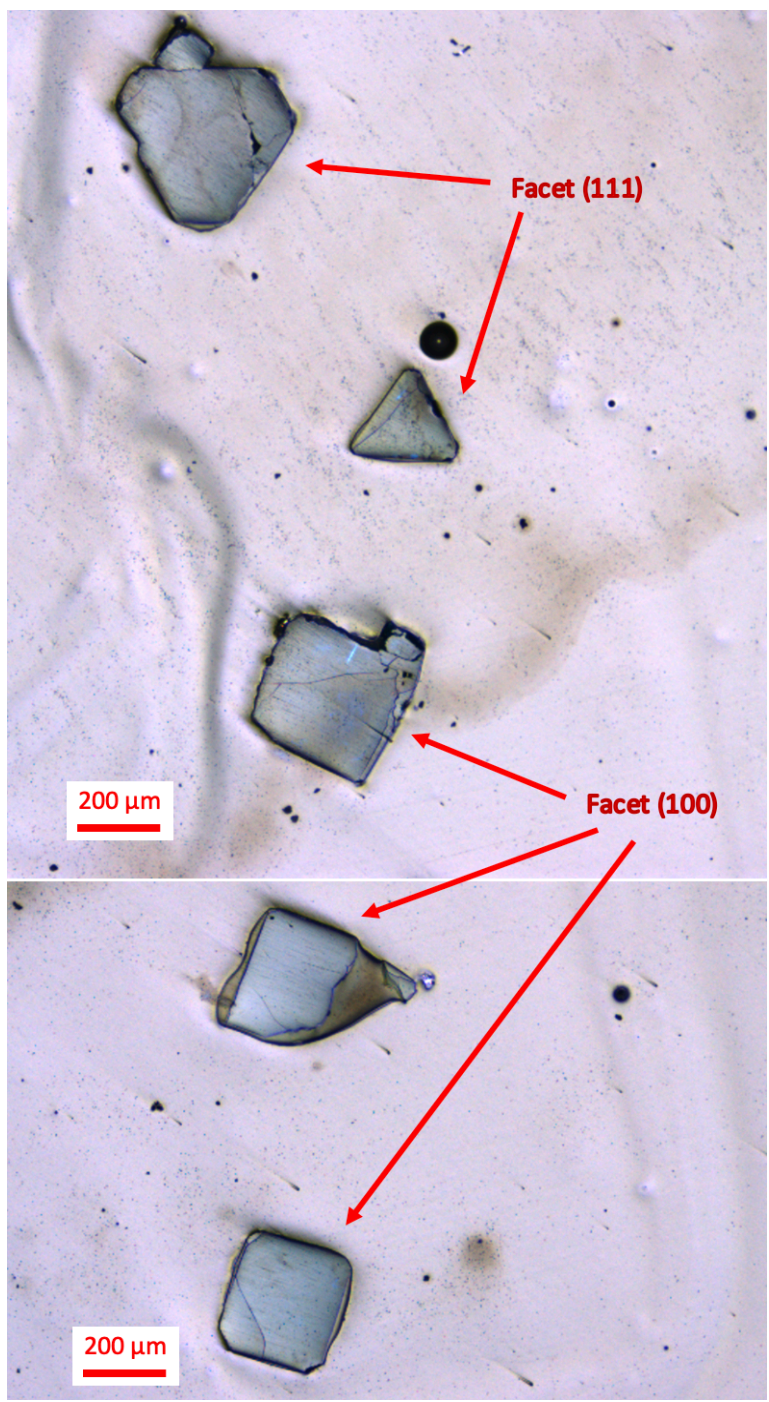

Figure S1. Single crystals of HKUST-1 oriented normal to the (111) and (100) facets, mounted on epoxy, carefully ground and polished to expose the smooth surfaces for FIB milling (see Figure S2).

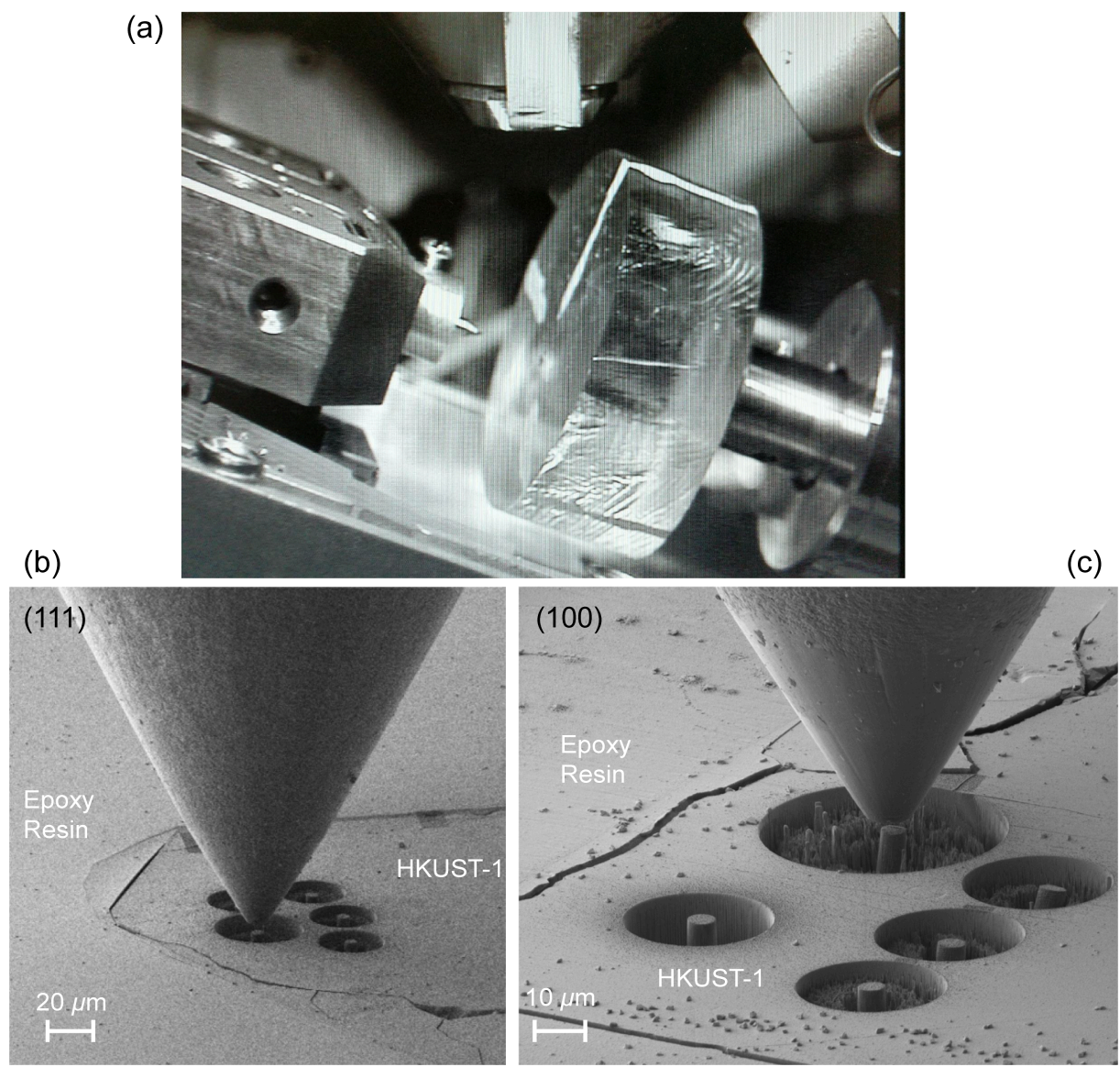

Figure S2. Optical and SEM mages of the specimens. (a) Optical image of the HKUST-1 specimen mounted on the Alemnis indenter positioned in a scanning electron microscope (SEM), in which the crystals were embedded on the surface of epoxy resin stub (thickness ~ 1 cm). *In situ* SEM images of the micropillars of HKUST-1 milled by focused ion beam on the (b) {111}- and (c) {100}-oriented crystallographic facets, respectively. Note that the two pillars located underneath the flat-punch indenter had just been tested.

Table S1. Diameters and heights of the micropillar specimens of HKUST-1.

| Crystal Orientation | Type of Test | Diameter, $D$ (μm) | Height, $L$ (μm) | $L/D$ ratio |
|---|---|---|---|---|
| (100) | Compression | 4.62 | 14.44 | 3.13 |
| | Compression | 4.92 | 15.47 | 3.14 |
| | Compression | 4.52 | 16.27 | 3.60 |
| | Splitting | 5.39 | 14.80 | 2.75 |
| | Splitting | 5.05 | 16.17 | 3.20 |
| (111) | Compression | 4.93 | 15.40 | 3.12 |
| | Compression | 4.76 | 14.00 | 2.94 |
| | Compression | 5.28 | 14.40 | 2.73 |
| | Splitting | 5.26 | 13.76 | 2.62 |
| | Splitting | 5.22 | 16.17 | 3.10 |

## 2. Buckling Strength / Pillar Stability / Euler's Critical Load

(a) Boundary Condition (BC) #1 – Bottom of the pillar is fixed but the top surface is free, the following equation gives the Euler's critical load ($P_E$):

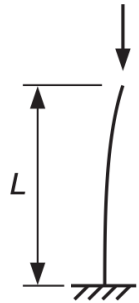

$$P_E = 0.25 \frac{\pi^2 EI}{L^2}$$

(b) Boundary Condition #2 – Bottom of the pillar is fixed and there is a mutual movement between the load cell and the top surface of the pillar, the Euler's critical load can be obtained using the following equation:

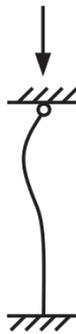

$$P_E = 2.05 \frac{\pi^2 EI}{L^2}$$

where $E$ is the Young's modulus; $I$ is the second moment of area (for a circular section, $I = \frac{\pi r^4}{4}$); $L$ is the height of the pillar.[2]

## 3. Uniaxial Micropillar Compression Videos and Micropillar Splitting Videos

Table S2. Video Files Recorded during the Pillar Micro-Compression and Splitting Tests

| Crystal Orientation | Video File Name | Compression (*Video frames accelerated ~15×*) | Splitting (*Video frames accelerated ~20×*) |
|---|---|---|---|
| (100) | (100) P1 comp | Pillar #1 | - |
|  | (100) P2 comp | Pillar #2 | - |
|  | (100) P3 comp | Pillar #3 | - |
|  | (100) P4 split | - | Pillar #4 |
|  | (100) P5 split | - | Pillar #5 |
| (111) | (111) P6 comp | Pillar #6 | - |
|  | (111) P7 comp | Pillar #7 | - |
|  | (111) P8 comp | Pillar #8 | - |
|  | (111) P9 split | - | Pillar #9 |
|  | (111) P10 split | - | Pillar #10 |

## 4. Instrumented Nanoindentation Results

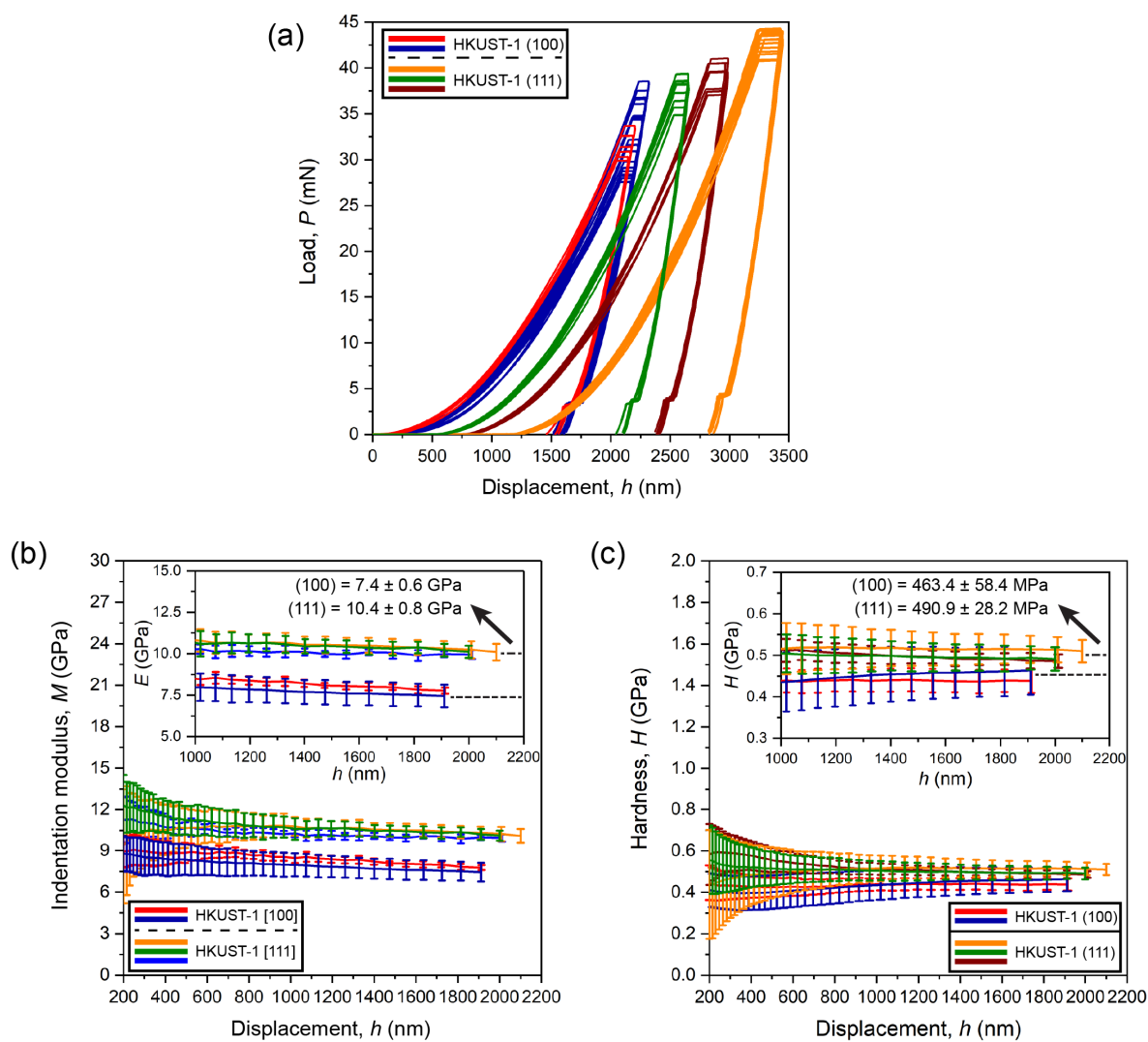

Figure S3. (a) Load-vs-displacement ($P$-$h$) curves from instrumented nanoindentation (MTS XP) of the two HKUST-1 crystal facets using a Berkovich tip. Each color represents a set of ~10 measurements. (b) Indentation modulus ($M = E/(1 - v_s^2)$),[3] and (c) hardness ($H$) data as a function of indentation depth, determined by continuous stiffness measurement (CSM) method. Insets show the average values derived from a surface penetration depth of 1000−2000 nm.

## 5. Yield Stress (*Y*) Determination

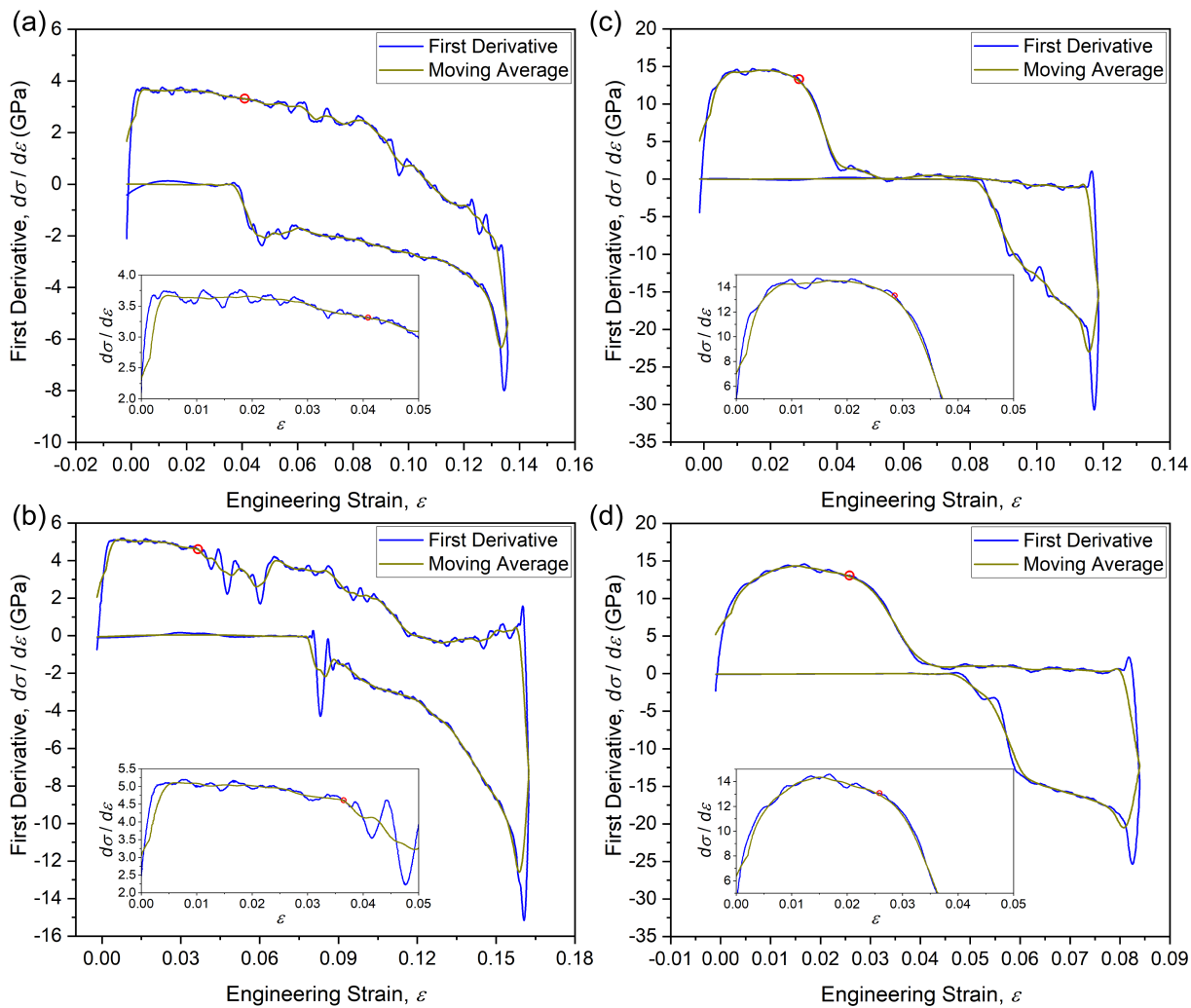

Figure S4. The first derivative of the stress-strain curve (dσ/dε) was smoothed out using the moving average method for the data obtained from the uniaxial compression test along the (a-b) [100]- and (c-d) [111]-axes. The yield points are highlighted in the red circles on the first-derivative curve, the views of which are enlarged in the corresponding insets.

To determine the yield point in a more rigorous way, we first differentiated the stress-strain (σ−ε) curve from the compression test with the Savitzky-Golay filter applied (4$^{th}$ order polynomial order, and points of window ≈ 3-7% of the total number of data points of the loading portion), and then smoothed out the first-derivative $\left(\frac{d\sigma}{d\varepsilon}\right)$ curve using the unweighted moving average method (the window length was set to be the same after the Savitzky-Golay filter used in the previous step) to highlight the fluctuations and trends. Subsequently, we applied 90% of the largest first-derivative value as the threshold and also indicator to identify the strain at the contact point and the yield point of the original stress-strain curve. The part of the stress-strain curve enclosed by these two points was regarded as the elastic regime, which was then linearly fitted to determine the value of Young's modulus ($E$). It is worth noting that the elastic limit and the yield point are typically very close, and hence herein we treated them as being coincident.

## 6. Gamma Coefficient and *H/E* vs. *Y/E* Ratios

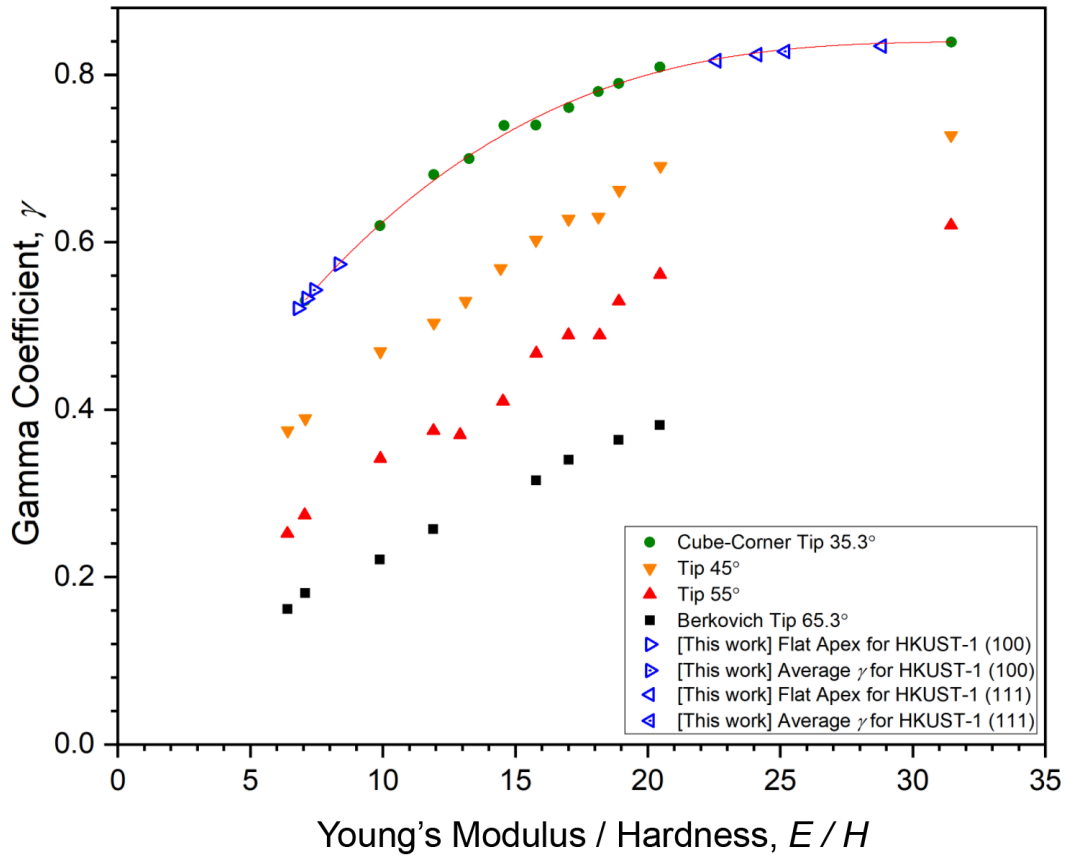

Figure S5. The gamma coefficient as a function of $E/H$ using the cube-corner indenter to implement the pillar splitting analysis of the two HKUST-1 facets, *viz.* the empty symbols in the figure. Moreover, the gamma coefficients acquired in this work are compared with the values of other materials using indenter tips of different geometries as reported by Ghidelli *et al.*[4] In Ghidelli's work, it should be noted that some of the data points were measured from pillar splitting technique while others were simulated by finite-element modelling (FEM). Adapted from ref. [4] with permission. Copyright (2017) The American Ceramic Society. (b) $H/E$ vs. $Y/E$ for the two HKUST-1 facets in comparison with the Tabor and Johnson relations.[5,6]

**References**

1   Wheeler, J. M. & Michler, J. Elevated temperature, nano-mechanical testing in situ in the scanning electron microscope. *Rev. Sci. Instrum.* **84** (2013).

2   Howatson, A. M., Lund, P. G. & Todd, J. D., Engineering Tables and Data (HLT). P. D. McFadden & P. J. Probert Smith, Eds.,  (Department of Engineering Science, University of Oxford, UK, 2009).

3   Tan, J. C., Merrill, C. A., Orton, J. B. & Cheetham, A. K. Anisotropic Mechanical Properties of Polymorphic Hybrid Inorganic-Organic Framework Materials with Different Dimensionalities. *Acta. Mater.* **57**, 3481-3496 (2009).

4   Ghidelli, M., Sebastiani, M., Johanns, K. E. & Pharr, G. M. Effects of indenter angle on micro-scale fracture toughness measurement by pillar splitting. *J. Am. Ceram. Soc.* **100**, 5731-5738 (2017).

5   Tabor, D. The physical meaning of indentation and scratch hardness. *Br. J. Appl. Phys.* **7**, 159-166 (1956).

6   Johnson, K. L. The Correlation of Indentation Experiments. *J. Mech. Phys. Solids* **18**, 115-126 (1970).